\documentclass[preprint2]{aastex}

\usepackage{amsmath}
\usepackage{graphicx}
\usepackage{epsfig,psfrag,epic,eepic}
\usepackage{textcomp}
\usepackage{times}
\usepackage{longtable}

\slugcomment{Submitted to the Astrophysical Journal}

\shorttitle{How Most Planets Form}
\shortauthors{Janson et al.}

\begin{document}

\title{How do Most Planets Form? -- Constraints on Disk Instability from Direct Imaging\altaffilmark{*}}

\author{Markus Janson\altaffilmark{1,5}, 
Mariangela Bonavita\altaffilmark{2}, 
Hubert Klahr\altaffilmark{3}, 
David Lafreni{\`e}re\altaffilmark{4}
}

\altaffiltext{*}{Based on archival data from the Gemini North telescope.}
\altaffiltext{1}{Princeton University, Princeton, USA}
\altaffiltext{2}{University of Toronto, Toronto, Canada}
\altaffiltext{3}{Max Planck Institute for Astronomy, Heidelberg, Germany}
\altaffiltext{4}{University of Montreal, Montreal, Canada}
\altaffiltext{5}{Hubble fellow}

\begin{abstract}\noindent
Core accretion and disk instability have traditionally been regarded as the two competing possible paths of planet formation. In recent years, evidence have accumulated in favor of core accretion as the dominant mode, at least for close-in planets. However, it might be hypothesized that a significant population of wide planets formed by disk instabilities could exist at large separations, forming an invisible majority. In previous work, we addressed this issue through a direct imaging survey of B2--A0-type stars, and concluded that $<$30\% of such stars form and retain planets and brown dwarfs through disk instability, leaving core accretion as the likely dominant mechanism. In this paper, we extend this analysis to FGKM-type stars by applying a similar analysis to the Gemini Deep Planet Survey (GDPS) sample. The results strengthen the conclusion that substellar companions formed and retained around their parent stars by disk instabilities are rare. Specifically, we find that the frequency of such companions is $<$8\% for FGKM-type stars under our most conservative assumptions, for an outer disk radius of 300~AU, at 99\% confidence. Furthermore, we find that the frequency is always $<$10\% at 99\% confidence independently of outer disk radius, for any radius from 5 to 500~AU. We also simulate migration at a wide range of rates, and find that the conclusions hold even if the companions move substantially after formation. Hence, core accretion remains the likely dominant formation mechanism for the total planet population, for every type of star from M-type through B-type.
\end{abstract}


\section{Introduction}
\label{s:introduction}

Direct imaging searches for exoplanets allow to constrain the population of wide and massive exoplanets. In a recent study \citep{janson2011b}, we used this fact to study the frequency of planets formed by disk instability \citep[e.g.][]{boss2003}, which are expected to have exactly these properties \citep[e.g.][]{rafikov2005,rafikov2007}. As we noted in that study, the population of close-in exoplanets discovered by indirect techniques very likely formed by core accretion, given lines of evidence such as the correlation of planet frequency with stellar metallicity \citep[e.g.][]{fischer2002} and the existence of the brown dwarf desert \citep[e.g.][]{grether2006}. In addition, studies of population distributions have demonstrated that most of the few wider-orbit exoplanets that have been directly imaged so far \citep[e.g.][]{marois2010,lagrange2010} are consistent with extensions of distributions defined by the closer-in sample \citep{crepp2011}. Hence, unless an additional wide population of planets formed by disk instability exists in even larger numbers than the presently known population, core accretion is the dominant mode planet formation. In our previous study, we searched for substellar companions around B-stars with direct imaging and compared the results to predictions of disk instability formation, in order to test the presence or absence of such a second population. We found that $<$30\% of the stars form and retain planets, brown dwarfs and very low-mass stars by this mechanism, at 99\% confidence. Meanwhile, the frequency of planets likely formed by core accretion is $\gg$34\% \citep{borucki2011}\footnote{Probably even $\gg$50\% \citep{mayor2011}.}, which indeed implies that this is the dominant formation mode for the total planet population.

Still, some uncertainties remain. For instance, the \citet{janson2011b} study was based on B-stars. Part of the reason for this was that the massive disks of B-stars should provide a very hospitable environment for disk instability formation. On the other hand, if for some unknown reason it should be the case that B-stars happen to be particularly poor hosts for this mechanism, then the comparison to the core accretion population around lower-mass stars could be irrelevant. Hence, in this new study, we extend our previous analysis to include FGKM-stars, by using archival data from the Gemini Deep Planet Survey \citep[GDPS,][]{lafreniere2007}. This data set has already been analyzed for providing statistical constraints on planet distributions \citep[e.g.][]{lafreniere2007,nielsen2010}, but not in the context of disk instability formation. Furthermore, an additional uncertainty in the original study was the outer disk radius, which was set to 300~AU based on the typical disk sizes in the few cases where they have been directly measured \citep[e.g.][]{pietu2007}. Here, we study what happens if the actual typical disk size is radically different, testing sizes all the way from 5 to 500~AU.

\section{Analysis}

Our procedure for determining formation limits from disk instabilities and evaluation of detectability with respect to the detection limits has already been described in \citet{janson2011b}, here we just summarize the main concepts, and introduce the features that are unique to this study. 

The formation limits are set by the fact that two criteria need to be simultaneously fulfilled for a planet to form by disk instability: The Toomre parameter \citep{toomre1981} needs to be small enough for an instability to occur, which translates to a minimum mass of the resulting planet (referred to as the 'Toomre criterion' here), and the cooling timescale needs to be short enough for fragmentation to occur \citep[e.g.][]{gammie2001}, which translates to a minimum semi-major axis of the resulting planet (called the 'cooling criterion' here). These criteria taken together delimit an allowed parameter range in mass versus semi-major axis space, within which planets and brown dwarfs can form. In addition, only companions with masses $<$100~$M_{\rm jup}$ are considered, and the outer disk radius is set to 300~AU by default (but we test radii all the way from 5 to 500 AU, as described below). An additional criterion that we consider here is an upper mass limit on the initial fragment from the local available mass, assuming a total disk mass of 0.5~$M_{\star}$. This limit was in principle already in place in our previous study, but did not have an impact in practice since it was $>$100~$M_{\rm jup}$ in the semi-major axis range where formation was otherwise allowed. In this study, where the stellar masses are lower, it does have some (though very minor) practical impact. The formation limits are generated individually based on the mass, metallicity, and initial luminosity of the star. In the majority of cases where the considered planets would orbit a single star or one component of a binary, the values of the parent star in question are used, but for the 14 cases where the planets would be circumbinary, the sum of masses and initial luminosities of the two components is used. The critical semi-major axes for binaries are determined following \citet{holman1999}. For the cases of unknown binary orbits, we set the semi-major axis to 1.31 times the projected separation following \citet{fischer2002}, and the eccentricity as the mean of the \citet{duquennoy1991} distribution, which is 0.36. These critical semi-major axes count effectively as additional formation limits in the cases where they apply, i.e. if a binary has a critical semi-major axis of 50~AU within which stable orbits are not possible, then in order for a simulated planet to be allowed to form, it must fulfill $a>50$~AU, in addition to the general formation criteria described above. Such limits were applied in the 19 cases where binarity sets constraints within 500~AU, the semi-major axis cut-offs are summarized in Table \ref{t:binaries}.

The term 'initial luminosity' in this context denotes the luminosity that the star would have had at the point when the hypothetical planet formed. This becomes a major uncertainty for stars of Sun-like mass and lower, because in the time frame that disk instabilities could be expected to occur, the star is still in its Hayashi contraction phase, which means that the luminosity evolves rapidly with age. In order to remain conservative and evaluate this uncertainty, we generate two sets of formation limits for each star, based on two extremes of initial luminosity, which we refer to as cases 'L1' and 'L2'. The two cases correspond to timescales which span an order of magnitude, basically corresponding to the full TTauri phase of star formation at ages 0.2~Myr to 2~Myr, and evolutionary models \citep{siess2000} are used to infer the luminosity from a given mass and age. Here, 'L1' corresponds to the earlier formation, and thus the higher luminosity, and 'L2' to the later formation. Stellar masses are inferred from the spectral types using the relation of \citet{kraus2007}. The metallicities are the same as in \citet{lafreniere2007} in the cases where they are listed there, and in some cases new metallicities from recent literature have been added \citep{alonso1996,ammons2006,paulson2006,mishenina2008,holmberg2009,jenkins2009,soubiran2010}. In the few remaining cases where the metallicity remains unknown, we have assumed Solar abundance. 

\begin{figure}[p]
\centering
\includegraphics[width=8cm]{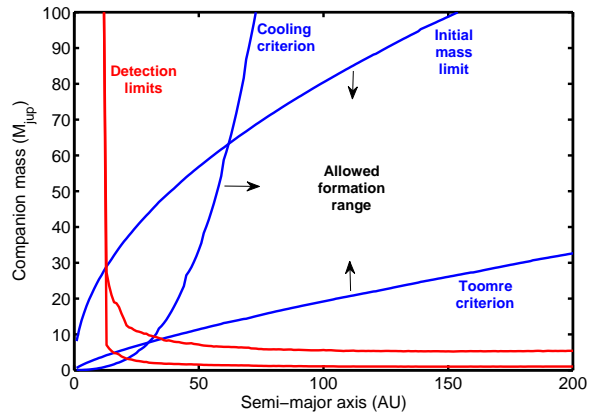}
\caption{Formation and detection limits for HIP 7235, which is a K0V-star with a wide possible age range of 100--1350~Myr. The formation limits correspond to the 'L1' case. The two detection limits correspond to the 'old' (upper) and 'young' (lower) cases, respectively. As most stars in the sample, the observations have a very high degree of completeness, regardless of age.}
\label{f:gdps5_l1}
\end{figure}

\begin{figure}[p]
\centering
\includegraphics[width=8cm]{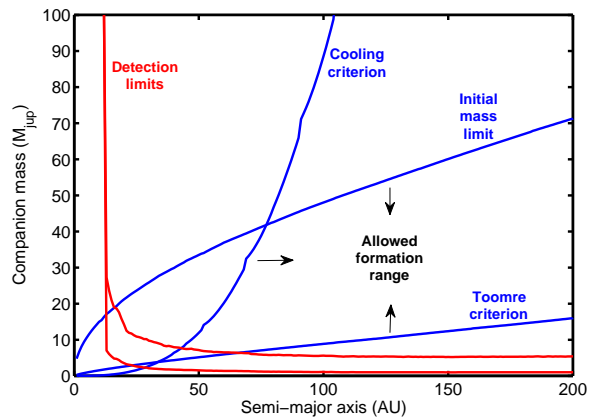}
\caption{Same as Fig. \ref{f:gdps5_l1}, but for the case of 'L2'. The completeness remains high, regardless of initial luminosity conditions.}
\label{f:gdps5_l2}
\end{figure}

In order to evaluate detectability within the formation limits, we use the MESS code \citep{bonavita2011} to generate 10000 orbits each in grid points of 5~AU spacing in semi-major axis and 1~$M_{\rm jup}$ spacing in mass. The eccentricities are uniformly distributed between 0.0 and 0.6 (the 'eccentric' case of our previous study). The projected separation and mass for each given orbit is compared to the detection limit from the images, and the fraction of orbits that are detectable at each given grid point is calculated. The total detection probability for the star is then calculated in two different ways: The 'uniform' case in which the planets are assumed to be uniformly distributed across the entire allowed formation space, and the conservative 'minimum' case in which the planets are assumed to be uniformly distributed along the minimum mass curve. In both of these cases, the semi-major axis distribution is assumed to be uniform. The detection limits are determined in the same way as in \citet{lafreniere2007}, using the 5$\sigma$ contrast limit translated into mass with COND-based models \citep{allard2001,baraffe2003} for temperatures below 1700~K and DUSTY-based models \citep{chabrier2000} for higher temperatures. The largest uncertainty in the detection limit determination is the age of the individual stars; indeed, for a couple of the stars in the GDPS sample, the age uncertainty spans more than an order of magnitude. In order to fully account for this uncertainty, we evaluate two extreme cases separately: One in which the lower age limits are used for all of the stars in the sample (referred to as the 'young' case), and one in which the upper age limits are used (the 'old' case). Examples of formation limits and how they relate to detection limits are shown in Fig. \ref{f:gdps5_l1} and Fig. \ref{f:gdps5_l2}.

The upper limit on the frequency of planets formed by disk instabilities, $f_{\rm max}$, is calculated for the full population in the same way as in \citet{janson2011b}, using binomial statistics and Bayes theorem. It is found that for an outer disk radius of 300~AU, less than $f_{\rm max} = 7.7$\% of the stars form and retain planets, brown dwarfs and very low-mass stars, at 99\% confidence, in the 'minimum', 'old', 'L1' case. This is the most conservative case, i.e. the one that yields the highest $f_{\rm max}$. The lowest $f_{\rm max}$ at 99\% confidence is acquired in the 'young', 'L2' case where $f_{\rm max} = 6.6$\% for both the 'uniform' and the 'minimum' distribution.

\begin{figure}[p]
\centering
\includegraphics[width=8cm]{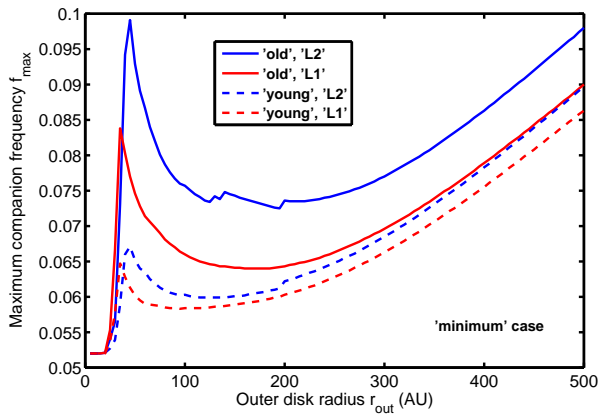}
\caption{Maximum frequency $f_{\rm max}$ as function of outer disk radius $r_{\rm out}$ (99\% confidence), for the 'minimum' distribution case. All 4 possible combinations of 'young'/'old' and 'L1/L2' are plotted. Even in the extreme case of 'L1' and 'old', the frequency never rises above 9.9\% within an outer disk radius of 500 AU.}
\label{f:minimumcase}
\end{figure}

\begin{figure}[p]
\centering
\includegraphics[width=8cm]{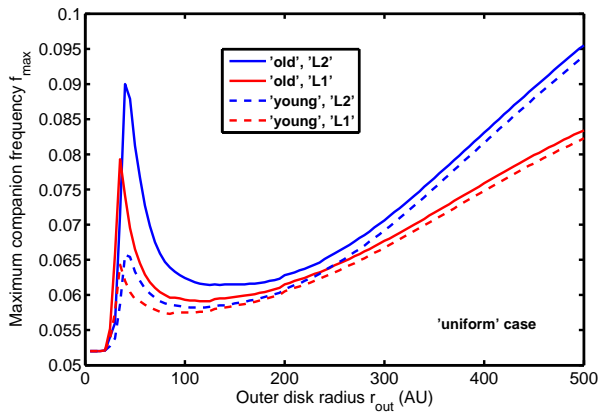}
\caption{Same as Fig. \ref{f:minimumcase}, but for the 'uniform' distribution case. Even in the extreme case of 'L1' and 'old', the frequency never rises above 9.6\% within an outer disk radius of 500 AU.}
\label{f:uniformcase}
\end{figure}

As a further step, we evaluate $f_{\rm max}$ as a function of the outer disk radius $r_{\rm out}$, for radii of 5~AU to 500~AU in steps of 5~AU. In the case where $r_{\rm out}$ is smaller than the inner edge of planet formation, the individual detection probability is simply counted as 1 for the purpose of calculating $f_{\rm max}$, since disk instability planets can't form under this circumstance in the first place (in other words, setting $f_{\rm max}$ to 1 implies that we are 100\% certain that no undetectable gravitational instability planets could have formed in these cases). The results of this test are shown in Fig. \ref{f:minimumcase} and Fig. \ref{f:uniformcase}. It can be seen that under any circumstance and for any outer disk radius below 500 AU, it is always the case that $f_{\rm max} < 9.9$\% at 99\% confidence. 

The curves in Fig. \ref{f:minimumcase} and Fig. \ref{f:uniformcase} have four distinct areas: For very small disks with outer radii of a few 10s of AU, planets simply cannot form by disk instabilities, so the maximum frequency is low. It then sharply peaks around $\sim$40~AU, where planets can form and simultaneously remain hidden from detection through projection effects in some instances. This is followed rapid decrease toward outer disk radii of $\sim$200~AU as a continuously decreasing fraction of the formed companions can hide through projection. A slower increase then follows for yet larger radii out to 500~AU, as an increasing fraction of the formed companions fall outside of the field of view.

\section{Discussion}

The result of this study significantly strengthens the conclusion of \citet{janson2011b} and extends it to FGKM-type stars: Core accretion, rather than disk instability, is likely the dominant form of planet formation. Below we provide a discussion of various aspect of this result, including several possible caveats that need to be taken into account.

\subsection{Planet frequencies}

The aspect of planet formation that we have mainly addressed in this paper is which mechanism constitutes the dominant mode for the total planet population. This includes every type of planet, i.e. both rocky and icy planets as well as gas giants. In a sense, this might be regarded as an inherently disfavourable comparison towards disk instability, since it a mechanism that is not expected to form any sub-giant planets in the first place, which are now known to be the dominant component of the core accretion population. However, such issues will always take place when two mechnisms are compared which have different predicted population distributions. Indeed, disc instability is expected to have favourable conditions to form very massive gaseous planets, which is difficult within the context of core accretion. It is fully conceivable a priori that disc instability could form a larger number of gaseous planets at large separations than core accretion forms rocky planets at small separations. Although the fact that this is apparently not the case is unsuprising, given the high efficiency of core accretion revealed by Kepler and radial velocity searches, it is the most relevant question that can be put forth with regards to the planet population at large, and one that has not been quantitatively addressed until now. As a side point, it should be noted that claims have recently been made that disk instability can form rocky planets, at least in the sense that the accumulation of the rocky body is kick-started in the core of a disk instability that subsequently fails to form a compact gaseous planet \citep{nayakshin2010}. We do not make any strong judgement on the relative credibility of this scenario here, other than to say that more theoretical work is probably required in order to distinguish it from the outcome of other mechanisms, in terms of observable quantities and distributions.

In any case, another relevant question that can be posed following this discussion is which mechanism is dominant for the formation of gas giant planets in particular. This question can also largely be addressed on the basis of our results. For instance, \citet{cumming2008} find a giant planet frequency of 10.5\% in their statistically complete sample, and predict a frequency of 17--20\% by extrapolation to larger separations. Even if we adopt the minimum value of 10.5\%, this is still higher than the upper limit of 9.9\% of disk instability planets that we find in our most conservative case. More recently, \citet{johnson2010} performed a similar study where they also study giant planet frequency as function of stellar mass. They find a lower value of 8.5\% in their statistically complete sample for FGK-stars. This should also increase if extrapolated to larger separations, but already the number of 8.5\% is higher than our upper limit for the vast majority of our explored conditions, and in fact, to get an upper limit lower than 8.5\% even in our most pathological case we only need to relax the confidence limit to 98\%. Hence, it holds true that core accretion is likely the dominant formation mechanism even for the restricted case of only gas giant planets, for Sun-like stars. The frequency of giant planets also increases strongly as a function of stellar mass in \citet{johnson2010}. However, this also means that it decreases to lower masses, and for M-stars it is only 3.3\%. Hence, for the specific case of gas giant formation around M-dwarfs, we cannot conclude which mechanism is dominant to any significant degree of certitude. We note that this distinction in stellar mass is not relevant in terms of comparing formation scenarios for the case of the total planet population. The planet frequency as function of mass is essentially flat in the Kepler sample \citep{borucki2011}. Although the frequency of gas giants decreases with stellar mass, the frequency of lower-mass planets actually increases toward lower stellar masses in their sample.

\subsection{Migration}

While the calculations up to this point have evaluated planets at the location where they formed, it is important to note that planets can migrate from their natal position to other places in the disk. However, while this mechanism can hide individual objects from detection, it cannot hide a dominant population formed by disk instability, as we will demonstrate below. In \citet{janson2011b}, we put forth two arguments to this effect: Firstly, that objects subject to type II migration are expected to grow rapidly in mass and would quickly hit the hydrogen burning limit and become stars, and secondly, that the migration rate would have to be extremely fine-tuned in order to hide such companions from both imaging and radial velocity surveys. Here, we quantify these arguments further. 

The planet-gas interaction that leads to type II migration in gaseous disks is expected to cause the planet to grow in mass. This is the mechanism by which \citet{mordasini2009} explain the existence of the majority of massive gas planets in a core accretion framework. These authors find a growth that is rapid and has a constant slope in logarithmic mass versus logarithmic semi-major axis, such that for every factor 2 that the planet travels in semi-major axis, it gains a factor 8.8 in mass. Hence, e.g. a 10~$M_{\rm jup}$ planet that migrates inwards from 50~AU will have a mass of 88~$M_{\rm jup}$ -- above the hydrogen burning limit -- by the time it reaches 25~AU. The increase in mass makes the object brighter, which makes them yet easier to detect, and furthermore, quickly brings them into the stellar regime where they are considered beyond the range of interest for our study.

Even if we ignore the growth issue, or assume that a mechanism exists that can cause planets to migrate without gaining mass, the migration rate (i.e. the distance within a gaseous disk that a planet can travel within the disk's life time) has to be highly fine-tuned in order to hide the planet from detection. If the planets travel at a low migration rate, they will be unable to reach the inner boundary of the direct imaging detection limits, so such an event can be excuded on the basis of our non-detections. If, on the other hand, they travel at high migration rate, then they would end up in the inner regions of the system and be detectable by radial velocity surveys (or plunge into the star). Hence, in this event, a dominant population of disk instability planets would constitute or strongly contaminate the observed radial velocity population. As we have discussed in Sect. \ref{s:introduction}, this would be inconsistent with the multiple lines of evidence suggesting that core accretion is the dominant mode of formation for the inner population, so such an event can also be excluded. This will be quantified in the remainder of this section.

The issue of exactly how planets migrate in the disk is hard to predict, as differences between recent modelling efforts show \citep[e.g.][]{baruteau2011,michael2011}. For instance, \citet{michael2011} simulate the first few thousand years of migration after a giant planet has formed. In one of their simulations, the planet migrates rapidly inwards by 6~AU in $\sim 10^3$ years and then halts, at least momentarily. If we interpret this in the sense that planets typically migrate $\sim$6~AU and then stop, then this motion is too small to hide any substantial number of objects inside the inner detection limit of imaging. If, on the other hand, we interpret it in the sense that 6~AU per 1000 years is the typical migration rate of planets, then in a typical disk lifetime of order $10^6$ years the distance traveled would be thousands of AU, such that all disk instability planets would be either hot Jupiters or consumed by their parent star. 

Given the uncertainties in quantifying the migration rate, we have performed an experiment to demonstrate that irrespective of this rate, it is exceedingly difficult to hide a population of disk instability companions through migration, given the highly confined parameter space that is available for this purpose. We do this by performing a series of similar simulations to our main simulation, but instead of evaluating the detectability of a given planet at its natal semi-major axis, we evaluate it at a smaller semi-major axis, given by a particular migration rate. For each simulation, we choose different migration rates, from 5 AU to 495 AU over the disk's lifetime, in steps of 5 AU (i.e., 99 simulations in total). In order to save computational resource, we only evaluate detectability in the 'minimum', 'L2', 'old' case, which is the most conservative set of assumptions possible (i.e., all other sets would yield tighter constraints on the resulting upper limit on companion frequency). If any companion falls within the innermost 5~AU after migration (including 0~AU), it is counted as detectable by radial velocity surveys. In the cases of circumbinary planets, we halt the migration at the inner stability boundary, inside of which there should be no disk material to migrate in. For each simulation, the upper limit on disk instability companion frequency $f_{\rm max}$ is evaluated for the full range of disk outer radii from 5 to 500~AU, as in our main simulation.

As a result, we have $f_{\rm max}$ as a function of both disk outer radius and migration rate in 99$\times$100 grid points. In 99.3\% of these points, $f_{\rm max} < 10$\% at 99\% confidence. In only one single case is $f_{\rm max} > 34$\%, which is our reference lower limit for core accretion. This is for the case of a disk radius of $\sim$30~AU and a migration rate of $\sim$20~AU per disk lifetime, in which a fairly large fraction of the objects formed can hide in the most elusive range between radial velocity and imaging detection. If we relax the confidence limit to 97\%, then $f_{\rm max} = 31$\%, i.e. under 34\% also in this case. Hence, we can conclude that very fine-tuned values for both the migration rate and the disk radius are simultaneously required in order to hide any significant number of targets, and even in that pathological case, we can still rule out a dominant disk instability population to a high degree of confidence. We also reiterate that this is for the 'minimum', 'L2', 'old' case, and that the upper limits would be even tighter for all the other cases. Finally, we can also note that if we e.g. relax the confidence limits to 95\%, then 99.6\% of the grid points give $f_{\rm max} < 8.5$\%, which is the lower frequency limit for gas giant planets around FGK-type stars found by \citet{johnson2010} as we discussed in the previous section, so we can also exclude disk instability as a dominant mechanism for the particular population of gas giant planets to a high degree of confidence, even in a migration framework.

\subsection{Other possible caveats}

Of course, given the complex and rather poorly understood nature of planet formation, there are other possible caveats that need to be considered. One frequent concern regards the uncertainty in the mass-luminosity models of young planets, in particular with regards to hot- versus cold-start models \citep[e.g.][]{fortney2008}. We discussed this issue already in our previous study \citep{janson2011b}, but to reiterate, cold-start models are constructed to apply specifically to the initial conditions in the core accretion scenario, and do not apply to disk instability scenarios. Since we are concerned with the detection of objects formed by disk instability specifically here, the cold-start models simply have no relevance. Apart from this point, we also showed in \citep{janson2011b} that even planets that presumably did form by core accretion do not fit the predicions by the cold-start models. Recent modelling efforts by e.g. \citet{spiegel2011} may be able to mitigate this latter point.

Another possibly relevant issue concerns uncertainties in the disk instability formation conditions. Our 'uniform' versus 'minimum' cases largely encapsulate this uncertainty, but one might imagine that the detailed physics of the formation process could alter the boundaries of the formation space. Indeed, recent work by \citet{kratter2011}, using a different approach for the cooling and fragmentation treatment, implies that formation might occur down to slightly smaller separations than previously believed. Furthermore, \citet{meru2011} have recently shown in simulations that a change in disk surface mass density profiles may affect the cooling criterion, which would also lead to formation at smaller semi-major axes. However, we note that even if the detailed boundaries are subject to uncertainties, it remains the case that there is an enormous parameter space further out around every star where such formation should be able to occur even more easily, yet it remains empty. Hence, from a population perspective, this issue is unlikely to matter significantly.

\subsection{Particular objects and classes}

Although disk instability does not seem to contribute substantially to the formation and retention of substellar companions, it is entirely plausible that individual objects may form in such a way, at the frequency level of a few percent or less. Probably, the best candidates for such companions, if they exist, are the $\sim$20--40~$M_{\rm jup}$ objects that have been found at $\sim$50--100~AU around Sun-like stars, such as GQ~Lup~B \citep{neuhauser2005,janson2006,mcelwain2007} or GJ~758~B \citep{thalmann2009,currie2010,janson2011a}. These could have formed in-situ by such a mechanism, whereas core accretion can most likely be excluded given the large masses \citep[however, regular binary formation also remains a viable option, see e.g.][]{bate2005}. Other interesting cases in this regard are those of very wide ($>$200~AU) and small planet-to-star mass ratio ($\sim$1\%) systems in young stellar associations like Upper Scorpius \citep{lafreniere2008,ireland2011,lafreniere2011}. These have too large separations and small masses to fulfill the Toomre criterion and form in-situ by gravitational instability. Hence, if we hypothesize that they did form by this mechanism, they would have to have migrated outwards. If they anyway had to migrate outwards, then core accretion remains an option to be considered as well, although the masses of the companions seem to be on the high end of what this mechanism can provide. In either case, for an object like HIP~78530~B \citep{lafreniere2011} in particular, it is unlikely that it could have migrated through interaction with a gaseous disk, since it is located at $\sim$700~AU and would therefore require a disk size of at least that order. Rather, it would have had to have been kicked out through dynamical interaction with another companion in the system. An interesting detail in this context is that all these very wide and low mass ratio systems have been detected in very young ($\sim$5~Myr) systems, despite the fact that they should be detectable also around older stars. This is consistent with predictions for dynamical interactions by e.g. \citet{veras2009}. In those simulations, a population of wide companions at hundreds of AU exist at ages of a few Myr, which have been kicked out by other companions in the system. These wide companions are then continuously ejected from the system through further interactions, and no longer exist to any significant extent at ages of $\sim$50~Myr and older. 

A related interesting point is the recent possible indication of large quantities of planetary-mass objects which may reside either at wide orbits or be free-floating, from gravitational microlensing \citep{sumi2011}. If real, we note that it is unlikely that such a population could have formed by disk instabilities, because the estimated characteristic mass of 1~$M_{\rm jup}$ is lower than the lowest-mass planets ($\sim$2--3~$M_{\rm jup}$) that can form even around M-type stars. The errors corresponding to a 95\% confidence level in the characteristic mass of the microlensing population do stretch up to $\sim$8~$M_{\rm jup}$, which would be a more reasonable match. However, with the much larger masses that would simultaneously be able to form through disk instability, it is not clear that the resulting mass function would be bottom-heavy enough to create the distinct population of microlensing event timescales observed by \citet{sumi2011}, whereas core accretion would naturally produce such a feature in the mass spectrum. Hence, on balance we do consider core accretion to be the more likely origin also for such a population of objects.

\section{Conclusions}

We have studied whether disk instability can form and retain a dominant population of substellar companions around FGKM-type stars, through comparison of theoretical model predictions to direct imaging data from the 85 stars in the GDPS survey \citep{lafreniere2007}. This work builds on a previous study of BA-type stars \citep{janson2011b} and extends it to a wider range of stellar masses. Our results confirm and strengthen our previous conclusion, with the finding that $<10$\% of the stars can form and retain such companions, at 99\% confidence, irrespective of outer disk radius for radii from 5 to 500~AU. Since the frequency of planets formed by core accretion is very likely $>34$\% \citep[e.g.][]{borucki2011}, it follows that core accretion is the most likely origin for the majority of the total planet population. This conclusion holds true at the 98\% or higher confidence level even if we restrict ourselves exclusively to the formation of giant planets, at least for FGK-type stars, where previous studies have found minimum giant planet frequencies of $\sim$9--10\% \citep{cumming2008,johnson2010}. In addition, we find that regardless of migration rate from 5~AU to 495~AU during the disk lifetime, it is not possible to hide a dominant population of disk instability objects simultaneously from direct imaging and radial velocity observations. 

\acknowledgements
The Gemini telescope is operated by the Association of Universities for Research in Astronomy, under a cooperative agreement with the NSF on behalf of the Gemini partnership. M.J. is funded by the Hubble fellowship.

{\it Facilities:} \facility{Gemini:Gillett (NIRI)}.

\begin{table}[p]
\caption{Binary limits}
\label{t:binaries}
\centering

\begin{tabular}{lccc}
\hline
\hline
GDPS ID & Other ID & $a_{\rm cs}$ (AU)\tablenotemark{a} & $a_{\rm cb}$ (AU)\tablenotemark{b} \\
\hline
8	&	HIP 11072	&	5.7	&	30.2	\\
11\tablenotemark{c}	&	HIP 13081	&	2.4	&	26.4	\\
14	&	HIP 14954	&	59.1	&	391.2	\\
23	&	HIP 30920	&	2.2	&	15.3	\\
28	&	HIP 44458	&	31.2	&	238.7	\\
38	&	HIP 54155	&	166.3	&	1264.2	\\
41	&	HIP 57494	&	16.4	&	125.3	\\
48	&	HIP 63742	&	0.3	&	3.1	\\
55	&	HIP 71631	&	4.9	&	59.9	\\
58	&	HIP 72567	&	28.5	&	195.2	\\
59	&	HIP 74045	&	5.1	&	33.7	\\
61	&	HIP 77408	&	185.3	&	1370.1	\\
65	&	HD 160934	&	2.0	&	15.5	\\
67	&	HIP 88848	&	0.9	&	10.5	\\
68	&	HIP 89005	&	18.4	&	136.1	\\
78	&	HIP 111449	&	85.9	&	628.3	\\
82	&	HIP 115147	&	135.9	&	975.0	\\
84	&	HIP 116384	&	7.0	&	54.0	\\
85	&	HIP 117410	&	11.2	&	85.8	\\
\hline
\end{tabular}
\tablenotetext{a}{Circumstellar outer semi-major axis limit.}
\tablenotetext{b}{Circumbinary inner semi-major axis limit.}
\tablenotetext{c}{The orbital parameters of the binary are from the preliminary spectroscopic fit of \citet{latham2002}. Furthermore, the system is a triple with an additional wide companion to this close binary pair. This sets a further constraint on the semi-major axis limit, such that circumbinary companions around the close pair can only reside between 26.4~AU and 284.6~AU}
\end{table}

\clearpage

\end{document}